# Pressure-stiffened Raman Phonons in Group III Nitrides

Gang Ouyang, [a, b] Chang Q. Sun [a, *] and Wei-Guang Zhu [a]

[a] *School of Electrical & Electronic Engineering, Nanyang Technological University, Singapore 639798, Singapore*

[b] *College of Physics and Information Science, Hunan Normal University, Changsha 410081, China*

**Abstract**

It has long been puzzling regarding the atomistic origin of the pressure-induced Raman phonon stiffening that generally follows a polynomial expression with coefficients needing physical indication. Here we show that an extension of the bond-order-length-strength (BOLS) correlation mechanism to the pressure domain has led to an analytical solution to connect the pressure-induced Raman phonon stiffening directly to the bonding identities of the specimen and the response of the bonding identities to the applied stimulus. It is found that the pressure-induced blue-shift of Raman phonons arises from the bond compression and energy storage exerted by the compressive stress. Agreement between predictions and experimental measurement leads to the detailed form for the polynomial coefficients, which offer an atomic understanding of the physical mechanism of the external pressure induced energy gain, thermally induced bond expansion as well as means of determining the mode atomic cohesive energy in a specimen.

# 1. Introduction

The group-III nitrides (GaN, AlN, and InN etc.) have attracted intensive attention due to their fundamental significance and technological applications such as short-wave-length light emitting devices, phonon transport and optoelectronic devices [1]. In general, the III-nitrides grown on substrate can induce biaxial internal strain and stress at the interface accompanied with deformation of the potentials for phonon and photon transportation. The local strain and stress act as trapping with large scattering cross section far away from resonant cases. The deformation potentials can be determined by combination of Raman spectroscopy under hydrostatic pressure and biaxial stress. The unit cell in a material would distort under high pressure, which lead to the zone-center phonon frequencies and the optical phonons stiffening. Characterizations of the group-III nitrides and other materials showed unusual physical and chemical properties under high pressure and high temperature [2-11]. generally, the frequencies of Raman mode increase with the applied pressure, which are well fitted using the quadratic function:

$$\begin{aligned}\omega(p) &= A\omega_0 + Bp + Cp^2, \text{ or} \\ &= \omega_0 + K^H p\end{aligned} \quad (1)$$

where $p$ is the pressure, $A$, $B$ and $C$ are the coefficients, $\omega_0$ is the frequency of a given Raman mode under zero pressure, $K^H = (\partial\omega/\partial p)_{p=0}$ is the hydrostatic linear pressure coefficient [10, 12, 13]. Evidently, the influence of external pressure plays a key role in the vibration behavior of the atoms in a specimen. Also, a homogeneous orthorhombic shear strain deformation would take place for the high-pressure approach [4]. Numerous theoretical models have been developed to elucidate the frequency-shift of Raman optical modes from the perspective of thermal expansion, interface contribution, and anharmonic phonon-decay, or their combinations [14-17]. At the present time, the high-pressure behavior of phonons dispersion relations and concomitant anharmonic effects in theoretical is studied using *ab initio* calculations based on a plane-wave pseudopotential method, local-density approximation (LDA), and generalized gradient approximation (GGA) within the density-function theory [18-20]. It would be more effective to connect the phonon properties directly to the lattice dynamics under the stimuli of pressure, temperature or their combination, which may result in the change of electronic structures. Unfortunately, little progress has been made on the atomistic understanding of the pressure-induced Raman shift.

Very recently, an approach of local bond average (LBA) [21] has been applied to model the thermal driven phonon softening in the temperature domain by relating the lattice dynamics to the response of bonding identities to the external stimulus such as temperature and coordination environment, which has been used successfully in the investigations of surface free energy of nanostructures [22-24]. In this paper, we extend the approach of LBA and the bond-order-length-strength (BOLS) correlation mechanism to studying the Raman phonon modes of GaN, AlN, and InN under hydrostatic pressure in terms of the functional dependence on the bonding identities (bond length, bond strength, and bond order). The developed theoretical methods show the frequency-shift of Raman optical phonons are dependent not only on the thermally induced bond vibration and bond expansion, but also on the work of external pressure that shortens and stiffens the bond.

**II. Principle**

For a given specimen no matter whether it is a crystal, non-crystal, or with defects or impurities, the nature and the total number of bonds remain unchanged unless phase transition occurs. However, the length and strength of all the involved bonds will response to the applied external stimulus. Therefore, we can focus on the response of the length and energy of all the bonds or an average of their representatives to the external stimulus if we consider only the property change of the specimen with the external stimulus. This approach of LBA collects statistic information from large number of bonds, which may represent the true situation of measurements and theoretical models. Differing from the volume partition approximation [25, 26] that focus on the absolute value of a certain quantity in the partitioned volume, the LBA approach is connected the relative change of a physical quantity with the applied external stimulus to the known bulk value, which focuses merely on the performance of the local representative atomic bonds disregarding the number of atomic bonds in the given specimen.

According to the LBA considerations [27], the total energy of a certain atom with z coordinates can be expressed as $E(d,T) = zE_b + \frac{zd^2u(r)}{2dr^2}\big|_d (r-d)^2 + \cdots$, where $d$, $T$ and $u(r)$ are the distance of two nearest atoms, the temperature and potential energy, respectively. Therefore, the phonon frequency can be obtained as a function of bond energy, atomic coordination, and bond length $\omega = z\left[\frac{d^2u(r)}{m^*dr^2}\big|_d\right]^{1/2} \propto \frac{z}{d}\left(\frac{E_b}{m^*}\right)^{1/2}$, where $m^*$ and $E_b$ represent, respectively, the reduced mass of the pair of bonding atoms, and the interatomic binding energy at 0K. The Raman phonon mode at 0K temperatures and zero pressure can be written as [27]

$$\omega(0) \propto \frac{z}{d_0}\left[\frac{E_B(0)}{m^*}\right]^{1/2} \quad (2)$$

where $d_0$ and $E_B(0)$ are the bond length and the atomic bond energy at zero temperature and zero pressure, respectively. The $\omega(0)$ is the specific Raman phonon frequency measured at zero temperature and zero pressure.

An extension of Eq. (2) to the pressure a temperature domain, we have the form for the temperature and pressure dependence of the Raman frequency,

$$\omega(p,T) \propto \frac{z}{d(p,T)}\left[\frac{E_B(p,T)}{m^*}\right]^{1/2} \propto \frac{[E_B(p,T)]^{1/2}}{d(p,T)} \quad (3)$$

with

$$\frac{d(p,T)}{d_0} = \left[1+\int_0^T \alpha_i(T)dT\right]\left[1+\int_0^P k_T dp\right] \quad (4)$$

$$\frac{E_B(p,T)}{E_B(0)} = 1 - \frac{1}{E_B(0)}\left[\int_0^T C(t)dt + \Delta E^p\right] \quad (5)$$

where $\alpha_i(T)$ and $k_T$ represent the thermal expansion coefficient and compressibility, respectively. $\Delta E^p$ is the distortion energy due to the applied pressure. Note that the $\Delta E^p$, showing in Figure 1, can be expressed as the integral area of $V_0 \triangle S$.

$$\Delta E^p = -\int_{V_0}^V p(v)dv \quad (6)$$

where $p$ and $V_0$ denote the external pressure, the volume of unit cell at zero pressure and zero temperature. The specific heat $C(t)$ from 0 K up to T or the conventionally termed specific internal energy follows the relations showing in the Ref [33].

Clearly, if the ambient temperature is much higher than the Debye temperature $\theta_D$, the specific heat approaches to a constant based on the Debye approximation. In reality, the specific heat follows neither the form of constant volume nor the form of constant pressure. For the first order approximation, we may take the Debye approximation as there is very tiny difference between the real situation and the ideal case [28].

Accordingly, the pressure-dependent Raman shift can be derived as follows,

$$\frac{\omega(p,T,z_i)}{\omega(0)} = \frac{d_0}{d_{(p,T)}}\left[\frac{E_B(p,T)}{E_B(0)}\right]^{1/2} \quad (7)$$

Noticeably, the integral of compressibility at constant temperature can be written as

$$\int_0^P k_T dp = -\int_0^P \frac{1}{V}\left(\frac{\partial V}{\partial p}\right)_T dp \quad (8)$$

The relationship between $V$ and $p$ can be obtained by the third-order Birch-Murnaghan isothermal equation of state [29, 30],

$$p(V) = \frac{3B_0}{2}\left(\xi^{-7/3} - \xi^{-5/3}\right)\left\{1 + \frac{3}{4}\left(B_0' - 4\right)\left(\xi^{-2/3} - 1\right)\right\}, \quad (9)$$

where $B_0$ is the static bulk modulus and $B_0'$ the first-order pressure derivative of the $B_0$, and the $\xi = V/V_0$ is the ratio of volume of unit cell addressed after and before being compressed.

Combining Eqs.(4)-(7), we can obtain the analytical form for the temperature- and pressure-dependent Raman shift,

$$\frac{\omega(p,T,z_i)}{\omega(0)} = \frac{\left\{1 - \frac{1}{E_B(0)}\left[\int_0^T C(t)dt + \int_{V_0}^V pdV\right]\right\}^{1/2}}{\left(1 + \int_0^T C(t)dt\right)\left(1 + \int_0^P k_T dp\right)} \quad (10)$$

Neglecting the temperature effect for system measured under the ambient temperature, Eq. (10) becomes the purely pressure-dependent,

$$\frac{\omega(p,T,z_i)}{\omega(0)} \approx \left[1 + p\Phi(\xi) + p^2\Phi^2(\xi)\right]\Theta(\xi) \tag{11}$$

where,

$$\Phi(\xi) = \left\{\frac{3B_0}{2}\left[\begin{array}{l}-\frac{7}{3}\xi^{-7/3} + \frac{5}{3}\xi^{-5/3} + \\ \frac{3}{4}(B_0' - 4)\left(-3\xi^{-3} + \frac{14}{3}\xi^{-7/3} - \frac{5}{3}\xi^{-5/3}\right)\end{array}\right]\right\}^{-1} \tag{12}$$

$$\Theta(\xi) = (1 - \Psi(\xi)/E_B(0))^{1/2} \tag{13}$$

with

$$\Psi(\xi) = \frac{3B_0V_0}{2}\left[\begin{array}{l}-\frac{3}{4}\xi^{-4/3} + \frac{3}{2}\xi^{-2/3} - \frac{3}{4} + \\ \frac{3}{4}(B_0' - 4)\left(-\frac{1}{2}\xi^{-2} + \frac{3}{2}\xi^{-4/3} - \frac{3}{2}\xi^{-2/3} + \frac{1}{2}\right)\end{array}\right] \tag{14}$$

Compared with the empirical polynomial expression, $\frac{\omega(p)}{\omega_0} = A + Bp + Cp^2$, the coefficient $A$, $B$ and $C$ are specified as, $A = \Theta(\xi)$, $B = \Phi(\xi)\Theta(\xi)$ and $C = \Phi(\xi)^2\Theta(\xi)$, with $\xi$ being the unique adjustable parameter.

### III. Results and Discussion

Using Eq.(10), we can calculate the pressure-induced Raman phonon stiffening of a specimen. Here we consider the Raman shift in group III nitrides with wurtzite structure as samples. The Debye temperature, the thermal expansion coefficient of *a*-axial and *c*-axial, and the single bond energy used in calculations are given in table 1. The intrinsic phonon frequencies $\omega(0)$ are obtained from the corresponding experimental results obtained under zero temperature and zero pressure. The thermal expansion coefficients of group III nitrides with hexagonal structures in *a*-axial and *c*-axial direction are employed the average value expressed as $\alpha = (2\alpha_a + \alpha_c)/3$ [31, 32].

Firstly, the pressure-induced external energy is calculated based on the Eq.(6). Figure 2(a) shows the distortion energy dependence on the $\xi$. Clearly, the energy increases nonlinearly with the pressure. The main reason is that the initial unit cell in the crystalline would be compressed upon pressure, and the value of $\xi$ becomes smaller. Also, the energy induced by pressure of GaN, AlN, and InN are almost the same. According to the Birch-Murnaghan isothermal equation of state, the $\xi$ values of the three nitrides are almost the same during external pressure process from 0 to 10 GPa. In addition, the square root of the mode cohesive energy could be different under the pressure. According to Eq.(5), the change of cohesive energy is dependent on the ambient temperature, and it leads to softening of red-shift of Raman phonons [33]. In particular, when the temperature is higher than the Debye temperature, the specific internal energy depends linearly on *T*. Secondly, the *d* (*p*, *T*) / $d_0$ ratio for group III nitrides at 300K, as shown in Figure 2(b), are different. Interestingly, the competition between the thermal expansion and the pressure-induced compression dominates the bond length. The bond length decreases with increasing the pressure at 300K, which implies the significance of pressure. The lattice constants *c/a* will differ remarkably from the original values and exhibit a drastic pressure variation.

Exceeding good agreement between theory and measurement of the typical Raman shifts has been realized as shown in Figure 3. The slow decrease of the Raman shift at 300K at zero pressure shows the temperature effect, as reported in ref [33]. Because the $\theta_D$ of AlN is higher than that of GaN and InN, the linearly variant of AlN can be extending to high temperature. However, the bond length $d$ will be shorter due to the high compressibility, and the $E_b(0)$ is also higher than that under zero pressure. According to the Birch-Murnaghan isothermal equation of state, the change of the volume of a unit cell of InN would be higher than GaN and AlN. So, the variation of bond length and the $E_b(0)$ for different vibration modes of InN are greater than that of GaN and AlN.

It is well known that the group III nitrides have several Raman active phonon modes at the Γ-point based on the group theory. For example, $E_1(LO)$, $E_1(TO)$, $A_1(LO)$, $A_1(TO)$, $E_2(high)$ and $E_2(low)$ modes are shown in GaN with wurtzite structure. The AlN wurtzite structure has also observed $A_1(TO)$, $E_1(TO)$, $E_2(high)$ and $A_1(LO)$ phonon modes in experiment [34]. While for the InN, long-wavelength $E_2$ and $A_1(LO)$ are concerned in wurtzite structure under hydrostatic pressure [10]. The $E_2(high)$ and $E_2(low)$ denote the bending mode of N-N and Ga-Ga, Al-Al and In-In. $A_1$ represents the axial direction and $E_1$ is the planar direction for vibration modes, respectively. Different atomic bond in a unit cell has different bonding energy. Since it is hardly possible to get the true value of mode cohesive energy, we have to investigate the pressure-induced Raman shift by fitting to the experimental data to gain information about the relative change. Each mode cohesive energies is just a portion of the specific bond energy, which are change under the external stimulus of pressure and temperature. Furthermore, according to Keating model, the two zone-center Raman frequencies satiated the relations [35-37],

$$\omega_{LO}^2(\Gamma) = \frac{1}{m^*}\left[4(\alpha+\beta-\sigma)+2\hat{f}/3\right] \qquad (15)$$

$$\omega_{TO}^2(\Gamma) = \frac{1}{m^*}\left[4(\alpha+\beta-\sigma)-\hat{f}/3\right] \qquad (16)$$

where $\alpha$ is the Keating force constant which describe the first-nearest-neighbor. $\beta$ is the force constant for the bond stretching between the concerned atom and the second-nearest-neighbors, and $\sigma$ the force constant for the bond-bending and stretch-stretch. The $\hat{f} = e^2 Z_B^2/(\varepsilon_0\varepsilon_\infty V)$ is Coulomb force constant, where $\varepsilon_0$ and $V$ are the vacuum dielectric constant and the volume of the primitive cell. The Raman modes mentioned above increase with compression of the bond length $d$ under high pressure, indicating that the $\alpha$ varies with the bond length in the $\sim d^n$ form with n being an adjustable parameter. The Keating's model is consistent with our theoretical consideration.

## IV. Summary

In summary, the physical mechanism of pressure-dependent blue-shift of the Raman optical modes has been addressed based on the thermally driven bond expansion and vibration, and pressure driven bond compression and stiffening. There is a competition between the two process of heating and compressing. Reproduction of the experimental measured $p$-dependent Raman shift of GaN, AlN, and InN at high pressure and 300K has been realized with deeper understanding of the atomistic origin and the physical mechanism behind the polynomial coefficients. The results evidence the essentiality to connect the macroscopically measurable quantities to the atomic bonding identities and the response of these bond identities to the pressure and temperature change.

**Acknowledgment**

The work reported in this paper is supported by the research grants ARC 04/06 and RG14/06 funded by MOE, Singapore, and by NSFC (No. 10772157), Hunan Normal University (070622), China.

**Figure captions**

Figure 1 Schematic illustration of the energy storage ($\Delta E^{\,p} = V_0 \Delta S$) induced by external compressive stress that causes bond energy elevation with $\Delta S$ being the area of integral from 1 to $\xi$.

Figure 2 Pressure-induced bond energy elevation ($\Delta E^{\,p}$) (a) and bond contraction and (b) for group III nitrides. The inset in (a) is the corresponding magnification area.

Figure 3 Agreement between theory (solid lines) and the measured pressure-dependent Raman frequency shift (scattered data) of (a) the $A_1$(LO) and the $E_2$ modes for GaN [2-4]; (b) the $E_1$(LO), $E_2$ and $A_1$(TO) for AlN [5-8]; (c) the $A_1$(LO) and $E_2$ for InN [9-10].

**Figure 1**

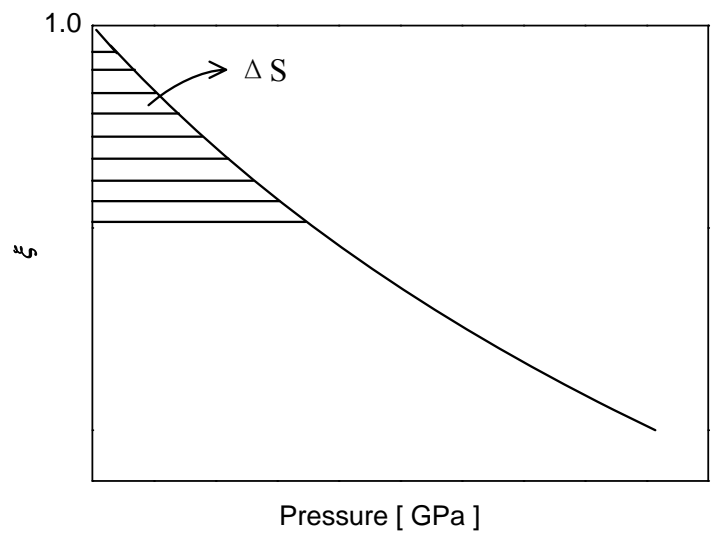

Pressure [ GPa ]

**Figure 2**

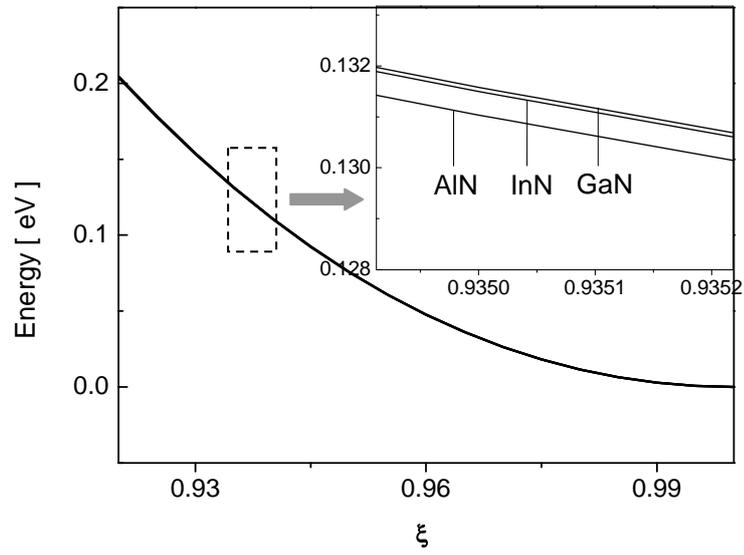

(a)

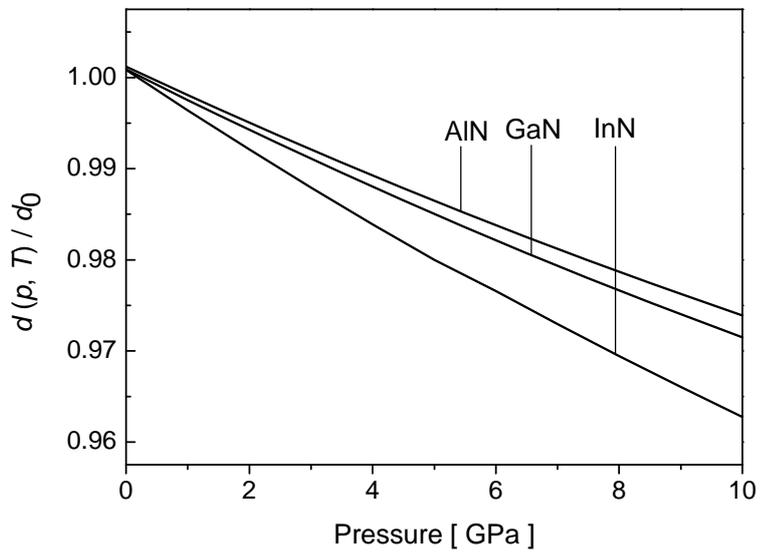

(b)

**Figure 3**

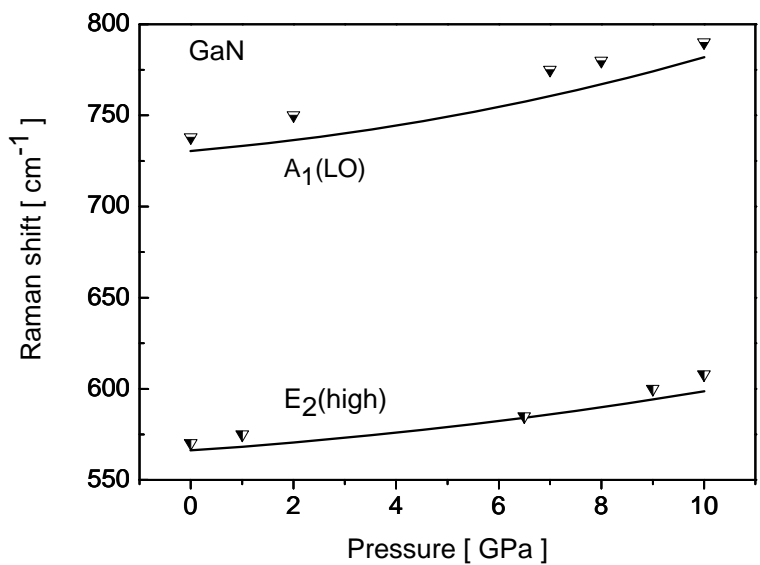

(a)

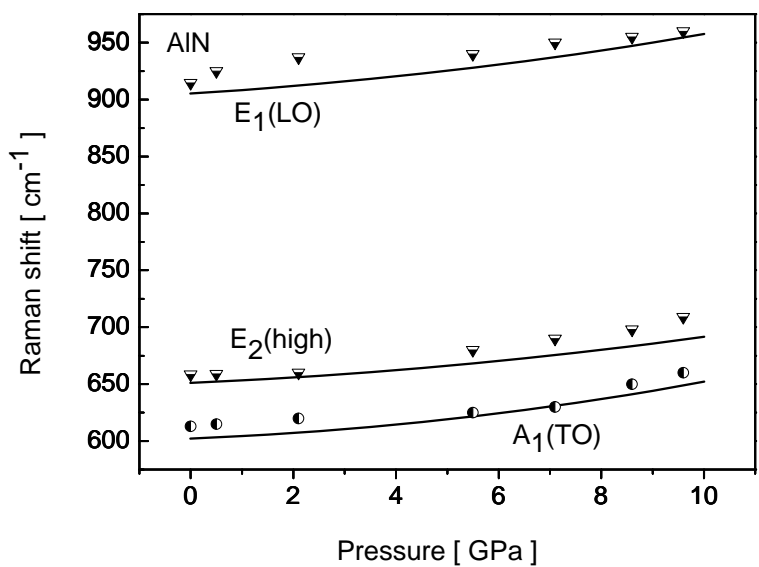

(b)

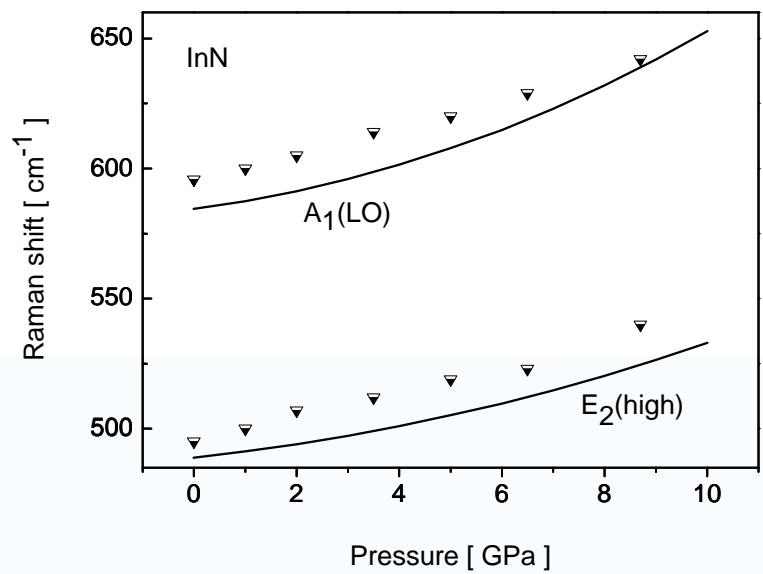

(c)

Table 1. The parameters used in the calculation of Group Ⅲ Nitride Crystals. The thermal expansion coefficients are listed as [33, 38, 39]: $\alpha_a = 0.232 \times 10^{-5} + 3.72 \times 10^{-9} T$ and $\alpha_c = 0.281 \times 10^{-5} + 1.783 \times 10^{-9} T$ for GaN, $\alpha_a = 0.3898 \times 10^{-5} + 1.5191 \times 10^{-9} T$ and $\alpha_c = 0.2925 \times 10^{-5} + 1.8698 \times 10^{-9} T$ for AlN, $\alpha_a = 0.2165 \times 10^{-5} + 2.60 \times 10^{-9} T$ and $\alpha_c = 0.2134 \times 10^{-5} + 6.08 \times 10^{-9} T$ for InN, respectively.

|  | $\theta_D$ (K) [33] | Raman mode [2, 5, 9] | $\omega_0$ (cm$^{-1}$) [2, 5, 9] | $E_b$(eV) [33] |
|---|---|---|---|---|
| GaN | 600 | A$_1$(LO) | 738 | 0.97 |
|  |  | E$_2$(high) | 570.2 | 1.44 |
| AlN | 1150 | E$_1$(LO) | 914.7 | 1.31 |
|  |  | E$_2$(high) | 658.6 | 1.13 |
|  |  | A$_1$(TO) | 613 | 0.71 |
| InN | 600 | A$_1$(LO) | 595.8 | 0.50 |
|  |  | E$_2$(high) | 495.1 | 0.76 |